\RequirePackage{ifpdf}
\ifpdf
	\documentclass[twocolumn, showkeys, showpacs, letterpaper, prb, pdftex]{revtex4}
\else
	\documentclass[twocolumn, showkeys, showpacs, letterpaper, prb, dvips]{revtex4}
\fi

\usepackage{amsmath}

\ifpdf
  \usepackage[pdftex]{graphicx}
  \usepackage{subfigure}
  \graphicspath{{pdf-figures/}}
  \usepackage[%
    pdftex,%
    pdftitle = {{Effect of grain shape on the agglomeration of polycrystalline thin films}},%
	pdfsubject = {{Grain shape can dramatically affect agglomeration of thin films. Even the slightest anisotropy of the interface energy can destabilize polygonal grains or, on the contrary, it can make them perfectly stable, independently of the cause of the faceting.}},%
	pdfauthor = {{Mathieu Bouville}},%
	pdfkeywords = {{grain-boundary grooving, dihedral angle, faceting, energy, theory, model}},%
  ]{hyperref}
\else
  \usepackage[dvips]{graphicx}
  \usepackage{subfigure}
  \usepackage[
	dvips,
    pdftitle = {{Effect of grain shape on the agglomeration of polycrystalline thin films}},%
	pdfsubject = {{Grain shape can dramatically affect agglomeration of thin films. Even the slightest anisotropy of the interface energy can destabilize polygonal grains or, on the contrary, it can make them perfectly stable, independently of the cause of the faceting.}},%
	pdfauthor = {{Mathieu Bouville}},%
	pdfkeywords = {{grain-boundary grooving, dihedral angle, faceting, energy, theory, model}},%
	letterpaper,%
]{hyperref}
\fi

\newcommand{\GB}{\ensuremath{_\mathrm{gb}}}
\newcommand{\I}{\ensuremath{_\mathrm{i}}}
\newcommand{\f}{\ensuremath{_0}}
\newcommand{\facet}{\ensuremath{_\mathrm{poly}}}
\newcommand{\iso}{\ensuremath{_\mathrm{circ}}}
\newcommand{\rb}[1]{\raisebox{-2.5ex}[0pt][0pt]{~$\Bigg\}\,$#1}~~~}

\begin{document}

\title[Grain shape and agglomeration of thin films]{Effect of grain shape on the agglomeration of polycrystalline thin films}

\author{Mathieu Bouville}
\email{m-bouville@imre.a-star.edu.sg}
\affiliation{Institute of Materials Research and Engineering, Singapore 117602}
\date{\today}

\begin{abstract}
The shape of the grains can dramatically affect the agglomeration of polycrystalline thin films by grain-boundary grooving. Anisotropy plays a central role in the stability against agglomeration of faceted films. Even a small difference between the interface energies of the facets can destabilize faceted grains or, on the contrary, it can make them perfectly stable at any thickness.
\end{abstract}


\maketitle

\begin{figure}
\centering
    \includegraphics[width=8.5cm]{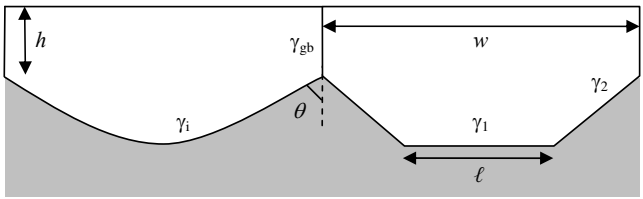}
\caption{\label{fig-schematic} A rounded grain and a polygonal grain. 
}
\end{figure}

Grain-boundary grooving occurs in all poly\-crystalline materials at the intersection between the grain-boundary and the interface or free surface. Grooving has been extensively studied theoretically.~\cite{Bailey-PPSB-50, Mullins-JAP-57, srol-JAP-86-I, miller-JMR-90, Genin-acta-93, thouless-acta-93, Ramasubramaniam-acta_mat-05, Xin-acta_mat-03, Zhang-acta_mat-04} It is a well known cause of agglomeration in thin films: if the groove becomes deep enough, the grains can become separate.~\cite{Srol-JOM-95}
Agglomeration therefore occurs when the height of the grain-boundary, $h$ in Fig.~\ref{fig-schematic}, reaches zero. The initial thickness of the film thus determines its stability.

At equilibrium, the shape of the interface must correspond to a constant chemical potential. Therefore, the interface is generally an arc of a circle at equilibrium. However, anisotropy of the interface energy, strain, or compositional inhomogeneities can affect the chemical potential and the interface may not be an arc of a circle in such cases.
For instance, the interface between a polycrystalline metal germano\-silicide thin film and the substrate is not an arc of a circle, unlike in metal silicides and germanides.~\cite{Aubry-JAP-02, Yao-05} 

Grain-boundary grooving in the context of faceted interfaces has attracted some attention.~\cite{Ramasubramaniam-acta_mat-05, Xin-acta_mat-03, Zhang-acta_mat-04} However, these works did not consider the case of thin films and the consequences on agglomeration of the shape of the interface.
In this Letter, we compare the agglomeration of thin films with rounded  and faceted interfaces, depicted in Fig.~\ref{fig-schematic} (the system studied is two-dimensional). By `rounded' we specifically mean that the interface is an arc of a circle.

The `volume' of a grain with an interface which is an arc of a circle is
\begin{equation*}
	V\iso = \left(\pi-2\theta-\sin2\theta\right) {r^2}/{2} + h\,w,
\end{equation*}
\noindent where $r$ is the radius of curvature of the interface. Since mass is conserved, $V\iso$ is also equal to its initial value, $w\,h\f$, where $h\f$ is the initial film thickness. As $\cos\theta=(w/2)/r$, one can write $V\iso$ as a function of $\theta$:
\begin{equation*}
	\frac{V\iso}{w^2}= \frac{h\f}{w} =\frac{\pi-2\theta-\sin2\theta}{8\cos^2\theta} + \frac{h}{w}.
	\label{def-V_iso}
\end{equation*}

Thinner films and larger grains are more prone to agglomeration. Under given experimental conditions and for a given grain size, there is a minimum thickness for the continuous film to be stable. We call $h\iso$ the minimum value of the initial film thickness to avoid agglomeration of a grain with an interface which is an arc of a circle,
\begin{equation}
	\frac{h\iso}{w}=\frac{\pi-2\theta-\sin2\theta}{8\cos^2\theta}.
	\label{def-h_iso}
\end{equation}
\noindent 
$h\iso$ depends on a single parameter: the groove angle $\theta$.

In the case of a polygonal grain, 
\noindent the minimum value of the initial film thickness to avoid agglomeration is
\begin{equation}
	\frac{h\facet}{w} = \frac{1}{4\tan\theta} \left[ 1 - \left(\frac{\ell}{w} \right)^{\!2} \right]\!,
	\label{def-h_facet}
\end{equation}
\noindent where the length $\ell$ is obtained by minimizing the total interface energy, $\gamma\GB\,h + \gamma_1\,\ell +\gamma_2 \, {(w-\ell)}/{\sin\theta}$, under constraint of mass conservation.\footnote{If annealing ends before the equilibrium grain shape has been reached, the groove has not reached its equilibrium depth yet. Equations~(\ref{def-h_iso}) and~(\ref{def-h_facet}) are thus upper bounds: thicker films are always stable but some thinner films may not agglomerate at short annealing times.}
It gives
\begin{equation}
	\ell = \lambda \, w,
	\label{def-ell}
\end{equation}
with 
\begin{equation}
	\lambda=\frac{1-\gamma_1/\gamma_2\,\sin\theta}{\cos\theta\,\cos\theta_0}
		\text{\quad and \quad}
	\cos\theta_0 = \frac{\gamma\GB}{2\gamma\I}.
	\label{def-R}
\end{equation}
\noindent $\lambda$ is the fraction of the grain width with energy $\gamma_1$. Note that $h\facet$ depends on three parameters: the groove angle $\theta$ and the energy ratios $\gamma\GB/(2\gamma\I)$ and $\gamma_1/\gamma_2$. 

In Eqs.~(\ref{def-h_facet}) and~(\ref{def-R}), $\theta$ is the angle between the facet and the grain-boundary. If the facet does not reach the grain-boundary (as shown for instance in Fig.~1(b) of Ref.~\onlinecite{Ramasubramaniam-acta_mat-05}) then $\theta$ is not the angle between the tangent to the interface and the grain-boundary but the angle between the facet and the grain-boundary. In what follows we assume that there is no such `rough' region or that it is small.

In the case of a rounded grain, the dihedral angle $\theta$ is set by the equilibrium of the grain-boundary and interface tensions~\cite{Bailey-PPSB-50, Mullins-JAP-57}
\begin{equation}
	\cos\theta=\frac{\gamma\GB}{2\gamma\I}.
	\label{def-theta}
\end{equation}
\noindent If the interface energy of polygonal grains is weakly anisotropic, the angle $\theta$ is set by the ratio of $\gamma\GB$ and $\gamma_2$ in the fashion of Eq.~(\ref{def-theta}). However, if the interface energy is strongly anisotropic the angle cannot change continuously, only some particular angles can exist.\cite{Wulff-1901, Ramasubramaniam-acta_mat-05, Xin-acta_mat-03, Zhang-acta_mat-04} Also, if the facet does not reach the groove root, its angle with respect to the grain-boundary cannot depend on the grain-boundary energy. In such cases, the angle~$\theta$ is not necessarily equal to the ratio of energies~$\theta_0$.
We will consider these two possibilities: (i) the angle is set by the ratio of energies, $\theta=\theta_0$, and (ii) the angle is independent of this ratio, $\theta \ne \theta_0$.

\begin{figure}
\centering
\includegraphics[width=8.5cm]{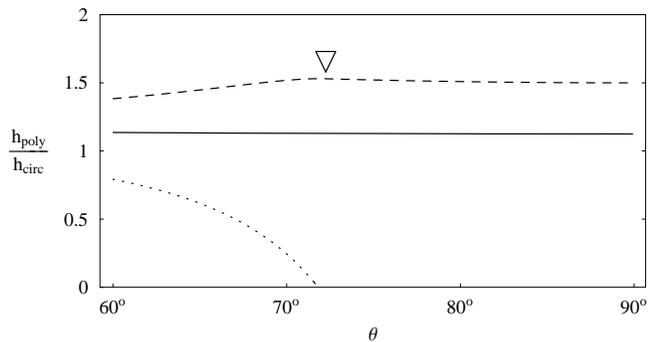}
\caption{\label{fig-ratio}The ratio of $h\facet$, the critical thickness for agglomeration of faceted grains, to $h\iso$, the critical thickness for agglomeration of rounded grains, as a function of the dihedral angle~$\theta$, for $\theta = \theta_0$. 
Dotted line: $\gamma_1=0.95\,\gamma_2$; solid line: $\gamma_1=\gamma_2$; dashed line: $\gamma_1=1.05\,\gamma_2$. At the arrow, $\ell=0$.}
\end{figure}

Figure~\ref{fig-ratio} shows $h\facet/h\iso$ as a function of the angle $\theta$ if $\theta=\theta_0$. If $\gamma_1 = \gamma_2$ (solid line), ${h\facet}/{h\iso} \approx {9}/{8} > 1$. A polygonal film is less stable than a rounded film (it needs to be thicker to avoid agglomeration). Since all interface energies $\gamma\I$, $\gamma_1$, and $\gamma_2$ are equal, this is a purely geometrical effect.
If $\gamma_1>\gamma_2$, polygonal grains are even more prone to agglomeration: a polygonal film needs to be about 40\% thicker than a rounded film to avoid agglomerating if $\gamma_1$ is larger than $\gamma_2$ by as little as 5\% (dashed line in Fig.~\ref{fig-ratio}). In this case, it is favorable for the grain to be `pointier' in order to reduce the amount of interface with the relatively large energy, $\gamma_1$. This favors agglomeration. 
If $\gamma_1<\gamma_2$, on the other hand, polygonal grains are more stable than rounded grains (dotted line in Fig.~\ref{fig-ratio}). In this case, energy is minimized by reducing the length of the interface with energy $\gamma_2$. It is therefore favorable for the grain to be `flatter', which tends to hinder agglomeration. 

\begin{figure}
\centering
\setlength{\unitlength}{1cm}
\begin{picture}(8.5,3.9)(.2,0)
\subfigure{
	\label{fig-ratio-08}
    \includegraphics[height=3.9cm]{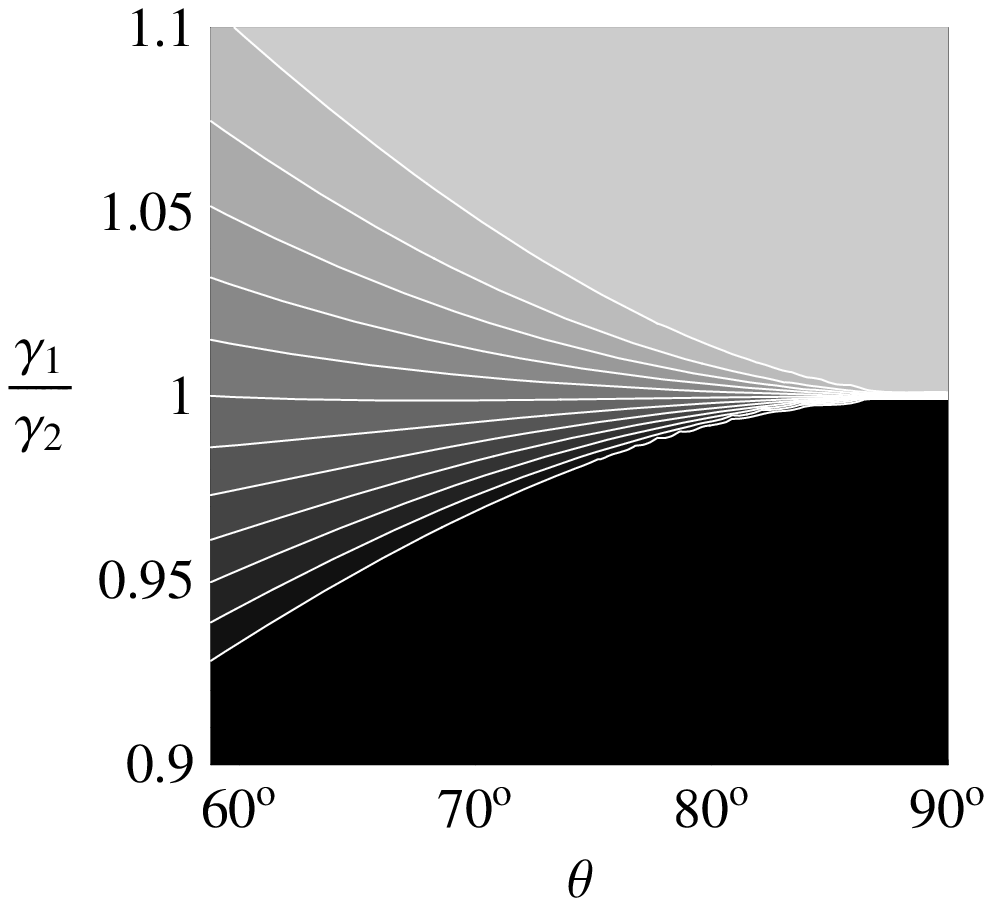}
	\put(-0.8,3.4){\bf(a)}
}\subfigure{
	\label{fig-ratio-12}
    \includegraphics[height=3.9cm]{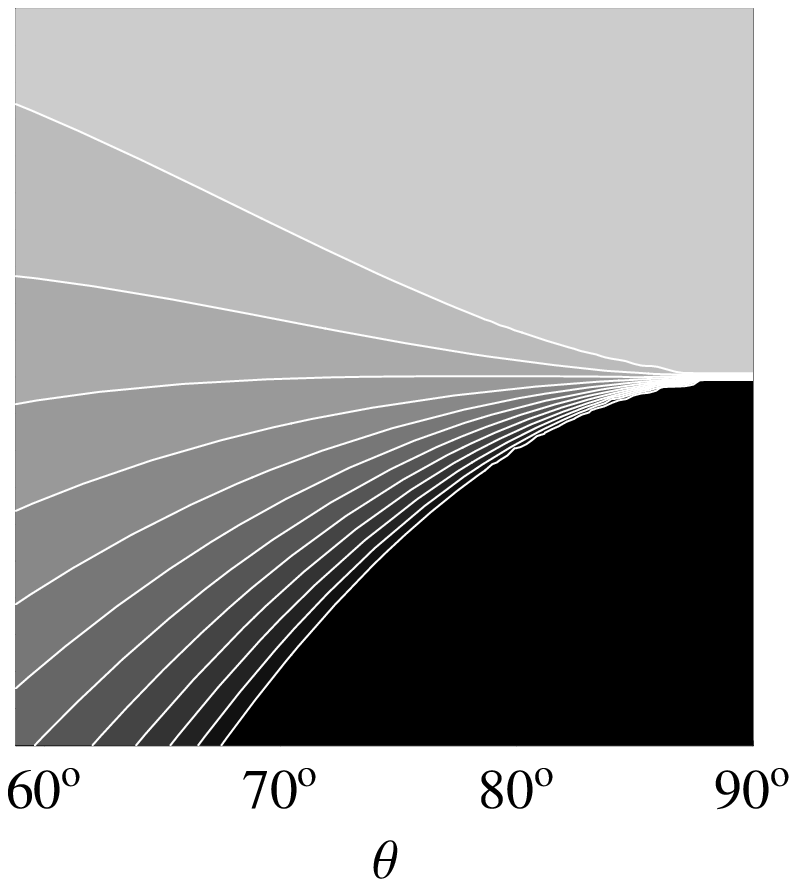}
	\put(-0.8,3.4){\bf(b)}
}\includegraphics[height=3.8cm]{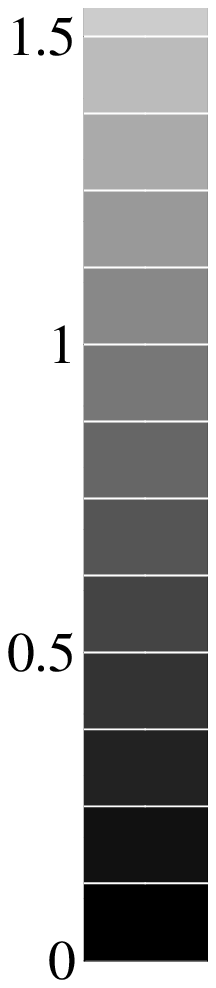}
\end{picture}
	\caption{\label{fig-ratio-contour}The ratio of the critical thickness for agglomeration of faceted grains to the critical thickness for agglomeration of rounded grains, $h\facet / h\iso$, as a function of the angle $\theta$ and of the ratio of interface energies $\gamma_1/\gamma_2$. (a): $\cos\theta_0 = 0.8\, \cos\theta$ and (b): $\cos\theta_0 = 1.2\, \cos\theta$.}
\end{figure}

If the dihedral angle is set by the anisotropy of the interface energy then the angle $\theta$ needs not be equal to the ratio of energies $\theta_0$. Figure~\ref{fig-ratio-contour} shows $h\facet/h\iso$ as a function of $\theta$ for different values of $\theta_0$.
The darker areas correspond to cases in which a polygonal film is more stable than a rounded film and in lighter areas an interface which is an arc of a circle is more stable than a faceted one. In the case of the black areas seen for small values of the $\gamma_1/\gamma_2$ ratio and large values of $\theta$, a polygonal film cannot agglomerate, however thin it may be (see below). 
The results for $\theta < \theta_0$ [Fig.~\ref{fig-ratio-08}] are similar to those for $\theta > \theta_0$, shown in Fig.~\ref{fig-ratio-12}. For a given value of $\theta$, the difference between $\theta$ and $\theta_0$ is less important than that between $\gamma_1$ and $\gamma_2$, i.e.\ the source of the faceting of the grains does not play a major role in the stability of the films. 

One should note that $\gamma_1 = \gamma_2$ is not incompatible with strong anisotropy. Anisotropy of the interface energy implies that $\gamma$ depends strongly on orientation and that facets which are actually observed are lower in energy than neighboring orientations. This does not imply that no two facets can have the same energy.

So far we assumed that the value of the length $\ell$ is set by energy minimization, as shown in Eq.~(\ref{def-ell}). However, this is not always true. 
For instance, it is geometrically impossible for $\ell$ to be lower than 0. If $({\gamma_1}/{\gamma_2}) \sin\theta > 1$,
\noindent then $\lambda < 0$ and Eq.~(\ref{def-ell}) leads to $\ell<0$. (Note that this is independent of the value of $\theta_0$.) In this case, $\ell=0$ and $h\facet=w/(4\tan\theta)$. This regime applies on the right of the arrow in Fig.~\ref{fig-ratio} and in the light areas of Fig.~\ref{fig-ratio-contour}. This is the case in which the polygonal grains are the least stable. Grains are then `pointy' as can be seen for instance in Fig.~2 of Ref.~\onlinecite{Jarmar-JAP-05}. One should note that this regime can exist only if $\gamma_1>\gamma_2$, in particular it cannot exist if $\gamma_1=\gamma_2$. The presence of such grains thus provides one with information on the $\gamma_1/\gamma_2$ ratio.

On the other hand, if energy minimization leads to $\ell > w$ (i.e.\ $\lambda > 1$) then $\ell=w$ and $h\facet = 0$. 
This occurs for the dotted line in Fig.~\ref{fig-ratio} for large $\theta$ and for the black areas in Fig.~\ref{fig-ratio-contour}. In such cases the film would not groove and {\itshape a fortiori} not agglomerate. Erbium disilicide films for instance do not exhibit any grooving. ErSi$_2$, which is hexagonal, grows epitaxially on Si(001) with its c-axis parallel to the substrate~\cite{tan-APL-06} but there probably is no epitaxial relationships for other directions due to the difference of crystal structure between film and substrate. Thus $\gamma_1$ is smaller than $\gamma_2$, which suppresses grooving.

For large values of $\theta$, either a polygonal film is always stable or it is much less stable than a rounded film. However, one should note that the latter instability may be only relative: both geometries may then be quite stable, with one even more stable than the other.

Table~\ref{table-R} shows that in all regimes, $h\facet$ is of the form $w/(4\tan\theta)f(\lambda)$. Consequently, $h\facet/h\iso$ is of the form
\begin{equation*}
	\frac{h\facet}{h\iso} = \frac{2\cos^3\theta}{\sin\theta(\pi-2\theta-\sin2\theta)} f(\lambda).
\end{equation*}
\noindent The fraction on the right-hand side depends weakly on~$\theta$:
\begin{equation*}
	\frac{2\cos^3\theta/\sin\theta}{\pi-2\theta-\sin2\theta} = \frac{3}{2} \left[ 1+ \frac{1}{5}\left(\frac{\pi}{2}-\theta \right)^2\right] + O\left(\frac{\pi}{2}-\theta \right)^4\!,
\end{equation*}
\noindent where $\theta$ is in radians, so that 
\begin{equation}
	\frac{h\facet}{h\iso} \approx \frac{3}{2} f(\lambda).
	\label{ratio-simple}
\end{equation}
\noindent If $\lambda < 1/\sqrt{3}$, Eq.~(\ref{ratio-simple}) leads to $h\facet > h\iso$: the polygonal grains are less stable than the rounded ones (the last column of Table~\ref{table-R} sums up the relative stabilities of polygonal and rounded grains depending on $\lambda$).

\begin{table}
\begin{tabular}{ccccc}
\hline
$\lambda$				& $\ell$		& $h\facet/w$					&shape		& stability\\
\hline
$\lambda \le 0$			& 0				& $\dfrac{1}{4\tan\theta}$		& `pointy'	& \rb{less}\rule[-1.5ex]{0ex}{5ex}\\
$0 < \lambda < 1/\sqrt{3}$& \rb{$w\,\lambda$}& \rb{$\dfrac{1-\lambda^2}{4\tan\theta}$}	&\rb{polygon}& \rule[-1.5ex]{0ex}{4ex}\\
$1/\sqrt{3} < \lambda < 1$& 				& 								& 			&  more\rule[-1.5ex]{0ex}{4ex}\\
$\lambda \ge 1$			& $w$			& $0$							& no groove &always\rule[-1.5ex]{0ex}{4ex}\\
\hline
\end{tabular}
\caption{\label{table-R} $\ell$, $h\facet$, the shape of the faceted grains, and whether faceted grains are more or less stable than rounded grains (or always stable, however thin), as functions of~$\lambda$.}
\end{table}

Whether polygonal grains are more or less stable depends only on $\lambda$, the ratio of $\ell$ over $w$, and not on the angle $\theta$ (of course $\lambda$ depends on $\theta$ but stability does not depend on $\theta$ \emph{directly}, only through $\lambda$). By simply measuring the length of the facet perpendicular to the grain-boundary compared to the grain width, one can infer whether faceting had a stabilizing or destabilizing effect.

Since $h\facet / h\iso$ depends on $\lambda$ only [Eq.~(\ref{ratio-simple})], we can use a Taylor expansion of $\lambda$ to quantify the effect of the parameters on $h\facet / h\iso$. 
Defining $\alpha$ by $\cos\theta_0 = \alpha \cos\theta$, to first order in $\alpha-1$ and $\gamma_1/\gamma_2-1$, one finds
\begin{equation}
	\lambda \approx \frac{1}{1+\sin\theta} \left[ 1-(\alpha-1)-\left(\frac{\gamma_1}{\gamma_2}-1\right)\frac{\sin\theta}{1-\sin\theta} \right]\!.
\end{equation}
\noindent Since $\sin\theta$ is close to 1, clearly $\gamma_1/\gamma_2$ plays a more important role than a difference between $\theta$ and $\theta_0$, as is noticeable in Fig.~\ref{fig-ratio-contour}.

To sum up, we showed that grain shape can affect agglomeration dramatically. For purely geometrical reasons, polygonal films generally need to be thicker than rounded ones to avoid agglomerating. If the interface energies of the facets ($\gamma_1$ and $\gamma_2$) are even slightly different, a polygonal film can be much less stable than a rounded one or, on the contrary, it may never groove and thus be stable at any thickness. This result depends weakly on the cause of the faceting of the grains.

\vspace{\parsep}
I wish to thank Dongzhi Chi and his group for useful discussions of their experimental results.

\bibliography{../NiPtSi-sim/silicides}

\begin{thebibliography}{15}
\expandafter\ifx\csname natexlab\endcsname\relax\def\natexlab#1{#1}\fi
\expandafter\ifx\csname bibnamefont\endcsname\relax
  \def\bibnamefont#1{#1}\fi
\expandafter\ifx\csname bibfnamefont\endcsname\relax
  \def\bibfnamefont#1{#1}\fi
\expandafter\ifx\csname citenamefont\endcsname\relax
  \def\citenamefont#1{#1}\fi
\expandafter\ifx\csname url\endcsname\relax
  \def\url#1{\texttt{#1}}\fi
\expandafter\ifx\csname urlprefix\endcsname\relax\def\urlprefix{URL }\fi
\providecommand{\bibinfo}[2]{#2}
\providecommand{\eprint}[2][]{\url{#2}}

\bibitem[{\citenamefont{Bailey and Watkins}(1950)}]{Bailey-PPSB-50}
\bibinfo{author}{\bibfnamefont{G.~L.~J.} \bibnamefont{Bailey}}
  \bibnamefont{and} \bibinfo{author}{\bibfnamefont{H.~C.}
  \bibnamefont{Watkins}}, \bibinfo{journal}{Proc.\ Phys.\ Soc.\ B}
  \textbf{\bibinfo{volume}{63}}, \bibinfo{pages}{350} (\bibinfo{year}{1950}).

\bibitem[{\citenamefont{Mullins}(1957)}]{Mullins-JAP-57}
\bibinfo{author}{\bibfnamefont{W.~W.} \bibnamefont{Mullins}},
  \bibinfo{journal}{J.~Appl.\ Phys.} \textbf{\bibinfo{volume}{28}},
  \bibinfo{pages}{333} (\bibinfo{year}{1957}).

\bibitem[{\citenamefont{Srolovitz and Safran}(1986)}]{srol-JAP-86-I}
\bibinfo{author}{\bibfnamefont{D.~J.} \bibnamefont{Srolovitz}}
  \bibnamefont{and} \bibinfo{author}{\bibfnamefont{S.~A.}
  \bibnamefont{Safran}}, \bibinfo{journal}{J.~Appl.\ Phys.}
  \textbf{\bibinfo{volume}{60}}, \bibinfo{pages}{247} (\bibinfo{year}{1986});
\textbf{\bibinfo{volume}{60}}, \bibinfo{pages}{255} (\bibinfo{year}{1986}).

\bibitem[{\citenamefont{Miller et~al.}(1990)\citenamefont{Miller, Lange, and
  Marshall}}]{miller-JMR-90}
\bibinfo{author}{\bibfnamefont{K.~T.} \bibnamefont{Miller}},
  \bibinfo{author}{\bibfnamefont{F.~F.} \bibnamefont{Lange}}, \bibnamefont{and}
  \bibinfo{author}{\bibfnamefont{D.~B.} \bibnamefont{Marshall}},
  \bibinfo{journal}{J.~Mater.\ Res.} \textbf{\bibinfo{volume}{5}},
  \bibinfo{pages}{151} (\bibinfo{year}{1990}).

\bibitem[{\citenamefont{Ramasubramaniam and
  Shenoy}(2005)}]{Ramasubramaniam-acta_mat-05}
\bibinfo{author}{\bibfnamefont{A.}~\bibnamefont{Ramasubramaniam}}
  \bibnamefont{and} \bibinfo{author}{\bibfnamefont{V.~B.}
  \bibnamefont{Shenoy}}, \bibinfo{journal}{Acta Mater.}
  \textbf{\bibinfo{volume}{53}}, \bibinfo{pages}{2943} (\bibinfo{year}{2005}).

\bibitem[{\citenamefont{Xin and Wong}(2003)}]{Xin-acta_mat-03}
\bibinfo{author}{\bibfnamefont{T.~H.} \bibnamefont{Xin}} \bibnamefont{and}
  \bibinfo{author}{\bibfnamefont{H.}~\bibnamefont{Wong}},
  \bibinfo{journal}{Acta Mater.} \textbf{\bibinfo{volume}{51}},
  \bibinfo{pages}{2305} (\bibinfo{year}{2003}).

\bibitem[{\citenamefont{Zhang et~al.}(2004)\citenamefont{Zhang, Sachenko, and
  Gladwell}}]{Zhang-acta_mat-04}
\bibinfo{author}{\bibfnamefont{W.}~\bibnamefont{Zhang}},
  \bibinfo{author}{\bibfnamefont{P.}~\bibnamefont{Sachenko}}, \bibnamefont{and}
  \bibinfo{author}{\bibfnamefont{I.}~\bibnamefont{Gladwell}},
  \bibinfo{journal}{Acta Mater.} \textbf{\bibinfo{volume}{52}},
  \bibinfo{pages}{107} (\bibinfo{year}{2004}).

\bibitem[{\citenamefont{G{\'e}nin et~al.}(1993)\citenamefont{G{\'e}nin,
  Mullins, and Wynblatt}}]{Genin-acta-93}
\bibinfo{author}{\bibfnamefont{F.~Y.} \bibnamefont{G{\'e}nin}},
  \bibinfo{author}{\bibfnamefont{W.~W.} \bibnamefont{Mullins}},
  \bibnamefont{and} \bibinfo{author}{\bibfnamefont{P.}~\bibnamefont{Wynblatt}},
  \bibinfo{journal}{Acta Metall.\ Mater.} \textbf{\bibinfo{volume}{41}},
  \bibinfo{pages}{3541} (\bibinfo{year}{1993}).

\bibitem[{\citenamefont{Thouless}(1993)}]{thouless-acta-93}
\bibinfo{author}{\bibfnamefont{M.~D.} \bibnamefont{Thouless}},
  \bibinfo{journal}{Acta Metall.\ Mater.} \textbf{\bibinfo{volume}{41}},
  \bibinfo{pages}{1057} (\bibinfo{year}{1993}).

\bibitem[{\citenamefont{Srolovitz and Goldiner}(1995)}]{Srol-JOM-95}
\bibinfo{author}{\bibfnamefont{D.~J.} \bibnamefont{Srolovitz}}
  \bibnamefont{and} \bibinfo{author}{\bibfnamefont{M.~G.}
  \bibnamefont{Goldiner}}, \bibinfo{journal}{JOM}
  \textbf{\bibinfo{volume}{47}}, \bibinfo{pages}{31} (\bibinfo{year}{1995}).

\bibitem[{\citenamefont{Aubry-Fortuna et~al.}(2002)\citenamefont{Aubry-Fortuna,
  Chaix-Pluchery, Fortuna, Hernandez, Campidelli, and Bensahel}}]{Aubry-JAP-02}
\bibinfo{author}{\bibfnamefont{V.}~\bibnamefont{Aubry-Fortuna}},
  \bibinfo{author}{\bibfnamefont{O.}~\bibnamefont{Chaix-Pluchery}},
  \bibinfo{author}{\bibfnamefont{F.}~\bibnamefont{Fortuna}},
  \bibinfo{author}{\bibfnamefont{C.}~\bibnamefont{Hernandez}},
  \bibinfo{author}{\bibfnamefont{Y.}~\bibnamefont{Campidelli}},
  \bibnamefont{and} \bibinfo{author}{\bibfnamefont{D.}~\bibnamefont{Bensahel}},
  \bibinfo{journal}{J.~Appl.\ Phys.} \textbf{\bibinfo{volume}{91}},
  \bibinfo{pages}{5468} (\bibinfo{year}{2002}).

\bibitem[{\citenamefont{Yao et~al.}()\citenamefont{Yao, Bouville, Chi,
  Tripathy, Sun, Pan, and Mangelinck}}]{Yao-05}
\bibinfo{author}{\bibfnamefont{H.~B.} \bibnamefont{Yao}},
  \bibinfo{author}{\bibfnamefont{M.}~\bibnamefont{Bouville}},
  \bibinfo{author}{\bibfnamefont{D.~Z.} \bibnamefont{Chi}},
  \bibinfo{author}{\bibfnamefont{S.}~\bibnamefont{Tripathy}},
  \bibinfo{author}{\bibfnamefont{H.~P.} \bibnamefont{Sun}},
  \bibinfo{author}{\bibfnamefont{X.~Q.} \bibnamefont{Pan}}, \bibnamefont{and}
  \bibinfo{author}{\bibfnamefont{D.}~\bibnamefont{Mangelinck}},
  \bibinfo{note}{{E}lectrochem.\ Solid-State Lett.\, accepted [preprint:
  cond-mat/0605451]}.

\bibitem[{\citenamefont{Wulff}(1901)}]{Wulff-1901}
\bibinfo{author}{\bibfnamefont{G.}~\bibnamefont{Wulff}},
  \bibinfo{journal}{Zeitschr.\ Kryst.\ Mineralogie}
  \textbf{\bibinfo{volume}{34}}, \bibinfo{pages}{449} (\bibinfo{year}{1901}).

\bibitem[{\citenamefont{Jarmar et~al.}(2005)\citenamefont{Jarmar, Ericson,
  Smith, Seger, and Zhang}}]{Jarmar-JAP-05}
\bibinfo{author}{\bibfnamefont{T.}~\bibnamefont{Jarmar}},
  \bibinfo{author}{\bibfnamefont{F.}~\bibnamefont{Ericson}},
  \bibinfo{author}{\bibfnamefont{U.}~\bibnamefont{Smith}},
  \bibinfo{author}{\bibfnamefont{J.}~\bibnamefont{Seger}}, \bibnamefont{and}
  \bibinfo{author}{\bibfnamefont{S.-L.} \bibnamefont{Zhang}},
  \bibinfo{journal}{J.~Appl.\ Phys.} \textbf{\bibinfo{volume}{98}},
  \bibinfo{pages}{053507} (\bibinfo{year}{2005}).

\bibitem[{\citenamefont{Tan et~al.}(2006)\citenamefont{Tan, Bouville, Chi, Pey,
  Lee, Srolovitz, and Tung}}]{tan-APL-06}
\bibinfo{author}{\bibfnamefont{E.~J.} \bibnamefont{Tan}},
  \bibinfo{author}{\bibfnamefont{M.}~\bibnamefont{Bouville}},
  \bibinfo{author}{\bibfnamefont{D.~Z.} \bibnamefont{Chi}},
  \bibinfo{author}{\bibfnamefont{K.~L.} \bibnamefont{Pey}},
  \bibinfo{author}{\bibfnamefont{P.~S.} \bibnamefont{Lee}},
  \bibinfo{author}{\bibfnamefont{D.~J.} \bibnamefont{Srolovitz}},
  \bibnamefont{and} \bibinfo{author}{\bibfnamefont{C.~H.} \bibnamefont{Tung}},
  \bibinfo{journal}{Appl.\ Phys.\ Lett.} \textbf{\bibinfo{volume}{88}},
  \bibinfo{pages}{021908} (\bibinfo{year}{2006}).

\end{thebibliography}
\bibliographystyle{apsrev}

\end{document}